# Data Siphoning Through Advanced Persistent Transmission Attacks At The Physical Layer

Alon Hillel-Tuch, *Member, IEEE*
Department of Computer Science
New York University Tandon School of Engineering
Brooklyn, NY 11201
ah5647@nyu.edu

*Abstract* — Data at the physical layer transmits via media such as copper cable, fiber optic, or wireless. Physical attack vectors exist that challenge data confidentiality and availability. Protocols and encryption standards help obfuscate but often cannot keep the data type and destination secure, with limited insight into confidentiality and integrity. We will investigate the feasibility of developing an awareness and integrity protocol to help mitigate physical side-channel attacks that lead to eavesdropping of data communication and denial-of-service.

## I. INTRODUCTION

Rao & Pfeffer describe the data stream when siphoning as a "continuous flow of personal information from the source to the destination"[1]. Covert data siphoning is not a new phenomenon, GCHQ in the UK has been siphoning data alongside other nations for several decades [2, 3]. In Operation Ivy Bells, the United States sent a specialized submarine (USS *Halibut*) into the Sea of Okhotsk. Divers installed a monitoring device around a Soviet undersea communications cable without compromising the cable's casing and recorded transmissions [4]. The project was eventually compromised when an NSA employee sold information to the KGB. What makes undersea data cables exceptionally intriguing is that we can consider them 'nation cables,' connecting entire countries and continents. The risk of critical compromise is real, "without interrupting transfer flow, you can read everything going on, [even] on an optical network." [5, 6].

To limit eavesdropping at the physical layer of the ISO-OSI model [7], we will need to explore how obfuscation and detection may mitigate data siphonage and enhance resilience to physical attacks. Typically, when an unauthorized party accesses a physical transmission, there is no state change to indicate a compromise to confidentiality. While encryption helps obfuscate data for unauthorized users, technological progress has shown encryption extends the time taken before an unauthorized party can access the data. Often through brute-force, social engineering, exploitable algorithmic flaws [8], [9], and physical limitation[10] [11] [12].

Eavesdropper exploitation appears to even be possible for person-in-the-middle attacks on certain quantum key distributions, leading to the ability to siphon data and decrypt, breaching data confidentiality as well as data integrity [13] [14] [15]. Air gapped systems with faraday cages have become susceptible to specific attack vectors, including mobile phones [16, 17]. There may even be implications around homomorphic encryption, which allows calculations on encrypted data without access to the key [18] [19]. One must assume the eventual decryption of any data; the constraint is time.

While some of these attack vectors are dependent on pre-installed malware, and primarily focus on how to relay accessed confidential data, there is still no mechanism to acknowledge siphoning. Integrity audits are not able to detect confidentiality breaches, as the integrity of the data remains intact. There appears to be no current mechanism for ascertaining whether a third party is privileged to the transmission. We will not know when a covert attack is underway, and often the attacks are supported by sophisticated actors, such as governmental organizations. We propose a protocol to help detect physical injections and increase network robustness.

The paper's organization is as follows; Section II describes the related work and limitations of current methods. In Section III, we describe the environment of persistent physical layer attacks and present a nodal framework. Section IV provides the motivating example behind our work. Section V describes the development of our proposed protocol. Section VI presents evidence from a simulation model. We conclude in Section VII and discuss future work.

## II. RELATED WORK AND CURRENT LIMITATIONS

Most research has covered attack vectors and how they could be implemented, rather than defensive strategies [20, 21]. Klimburg's research identified that Western democracies have a high dependency on volunteerism when looking at public policy cooperation for APT defense, unlike counterparts in countries such as China and Russia, which have strong public-private cooperation, either mandated or participatory [22].

Geers' research focused not on systemic defense but deterrence using the analogy of an atomic bomb in traditional military deterrence tactics, calling it Mutually Assured Disruption (MAD) [23]. Uma et al. provided a survey on cyber-attacks, including nation-state, and their classifications [24]. They fail to recognize that physical layer attacks are possible beyond Wireless Sensor Networks or Mobile ad-hoc Networks, a commonly neglected vector in nation threat models. Research surveys on attacks tend to consider APT attacks as a generic classification of government-level cyber warfare [25] or network-level attacks targeting IoT devices/sensors instead of transmission lines [26] [27] [28] [29].

There are some parallels between our research and those made on power grid vulnerabilities. In an analysis of data injection attacks on power grids Kim et al. suggests identifying a subset of measurements to be selected and made immune to threats[30]. The structure of electric grids presents a similar high-complexity combinatorial problem of selecting subsets as our nodal network. The proposition is to use a fast greedy algorithm that selects a subset of measurements. We will build from this research and apply a fast greedy algorithm to identify a subset of data packets to protect.

To model the threat, we need to bifurcate local and national level threats. We further modify the discrete systems approach taken by Wardens Five Concentric Ring theory [31], connected to cyber threats by Ogîgău-Neamțiu et al. [32]. The original model describes an enemy force and the various attack vectors upon it. Targets closer to the center of the rings have a more significant effect on the enemy (the core is leadership). The model will instead consider the enemy force as the attacker, not the defender. We will isolate state actors' risk and classify national cyberinfrastructure and residential, commercial, industrial. We will be modifying 'fielded military' to become 'defense mechanism'.

TABLE 1 – MODIFIED WARDENS CONCENTRIC RING THEORY ANALYSIS

| | Service Denial and Degradation |
|---|---|
| Leadership | Ability to coordinate and delegate severely impacted. Isolated communication networks may function, particularly those independent of the affected service denial (i.e., satellite communication). |
| System Essentials | Nonexistent to severely strained. |
| Infrastructure | Nonexistent to severely strained. |
| Population | Unable to access information, seek assistance, report damage |
| Defense Mechanism | With no active communication line, most defense mechanisms will be inactive. The only appropriate defense mechanism would be a second communication line for infrastructure to use or a rapid-response restoration. |

Any attack that can selectively deny or degrade communications service will directly affect all five rings. As communication degrades, defense mechanisms, population, infrastructure, system essentials, and leadership break down. Under the primary CIA triad lens, service denial and degradation remain a real risk regardless of channel security, such as encryption, to protect Confidentiality and Integrity. Mitigation of Disruption as a threat type is in most cases mutually exclusive from Disclosure, Deception, and Usurpation, affecting Availability.

Prior research has explored solutions to obfuscate the physical location and structure of critical networks [33] as well as protect optical transmission and cables [34] [35] [36]. While we recognize that specific systems, such as satellite communication, may still function if a nation cable is physically severed, it is essential to note that those systems cannot handle the throughput demanded by a functioning infrastructure. Even $1/10,000^{th}$ of the throughput of the denied service would overwhelm existing satellite communication infrastructures. That said, recent commercial developments such as Starlink may help decentralize and offload ground-based infrastructure [37]. While specific environmental impacts have been identified [38], there are legitimate concerns on the lack of political research into the potential monopolization of a global data infrastructure by a commercial non-governmental entity, and the inherent risk around data centralization and data sovereignty.

### III. NODAL FRAMEWORK

Public data on attacks and infrastructure outages is restricted, with many government agencies globally deliberately limiting information access. For example, The US Department of Homeland Security deemed public data on infrastructure outages to be a national security risk, and as such, filed with the FCC requests to lock up public data, as they stated, "even a single event may present a grave risk to the infrastructure" [39].

Our proposed model uses the concept of an idempotency key to identify duplicate and original data (not subsequent retries) for validation. The proposal aims to create multiple pathways for data transmission and transmit those packets marked as critical through the pathways in parallel, with a fast greedy algorithm selecting a subset of non-critical data for multi-path transmission to handle injection and data corruption detection.

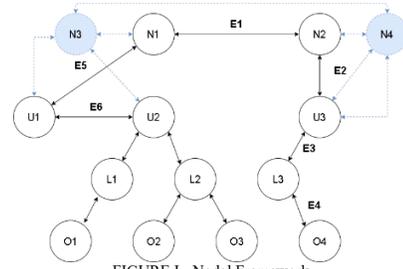

FIGURE I - Nodal Framework

Nation level nodes ('N') handle top-level communication, outer level nodes ('O') are the end-user systems. A message from O1 to O4 has only one path (O1->L1->U2->U1->N1->N2->U3->L3->O4). We can see that terminating the N1<->N2 edge (E1) would be the greatest disruption to all nodes, assuming randomly distributed transmission between all nodes. A redundant N level edge (N3<->N4) would increase robustness. The path is not guaranteed the most efficient node-hop – this is deliberate. Trace-routing is supported and presented similar to IEPM-BW's route trace paths[40].

### IV. MOTIVATING EXAMPLE

The main problem is physical link-layer attacks that manipulate and deny/sever service outlined by our threat model[41]. This attack type has the most significant and immediate impact in a nation attack with a near equivalent level of effort than other comparable attacks. While we hypothesize that attacks mainly focus on nation cables and extensive commercial communication infrastructure, the hypothesis is that by creating a distributed communication network based on priority, we can provide near synchronous data redundancy and an integrity mechanism to reduce threat risk measurably.

### V. DEVELOPMENT OF PROPOSED PROTOCOL

We will be using our previously run simulation using the NNAS address convention and a Mixed-ANOVA model (N=100, dF=99). For detailed results, please see *Nodal Framework Testing Environment* [41]. Typically, when assessing the performance of a network, there are seven (7) critical network metrics: bandwidth usage, throughput, latency, packet loss, retransmission, availability, connectivity [19] [42]. We will not be looking at bandwidth, latency, and throughput. Our measurement will be around packet loss, retransmission, availability, and connectivity. Also, we will measure corruption, which is a counter of compromised data.

## VI. EMPIRICAL EVIDENCE

Our simulation generates random attacks and measures our proposed distributed system protocol's efficacy. The simulation model records the metrics and stores them in a CSV file for analysis.

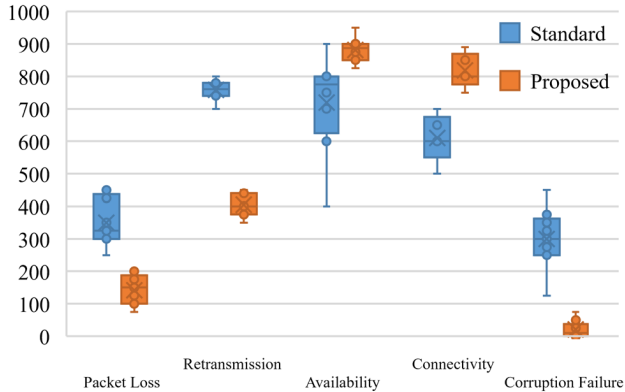

FIGURE II- BOX WHISKER PROTOCOL COMPARISON

We measure significant reduction in packet loss (Δ-200, 50%), retransmission (Δ-300, 42%), and undetected corruption (Δ-275, 91.6%). Connectivity and availability were enhanced. We notice under the standard model that availability and corruption detection has a high degree of variance; this is most likely due to the binary nature of packet loss and the proposed model's inherent double redundancy in communication. As discussed in this paper, corruption detection is only possible if we have a verified copy and hash of the original transmission.

## VII. CONCLUSIONS AND FUTURE WORK

There are two main variables worth specific consideration, and that is availability and corruption. Packet Loss, Retransmission, and Connectivity all correlate with these two measures. High availability would reduce packet loss, retransmission, and increase connectivity.

The proposed protocol significantly improves the ability to detect corruption and perform error correction, a measured 91.6% improvement. However, there is significant overhead in corruption detection. While we tried to mitigate the overhead by applying a fast greedy algorithm to create a subset of the dataset transmitted, that subset is still additional work. Further research into the overhead costs is warranted. Furthermore, as per protocol, critical data packets have a parallelized transmission across unused network nodes.

The current simulation does not account for latency and assumes a zero-latency system (not realistic). We can theorize that the transmission latency would be equal to the highest latency in the network between the two identical transmissions. One transmission most likely goes via the most (best-effort) efficient path dictated by existing network structures. The parallel transmission's only criteria are not to use any previously used path, warranting further analysis on throughput and network load.